%% file: Paper.tex
\documentclass[usenatbib,usegraphicx,useAMS]{mn2e}
\usepackage{amsmath}
\usepackage{amssymb}

\input{journal_abbr}  
\input{unit_abbr} 

\title{The Fate of Former Companions to Hypervelocity Stars Originating at
the Galactic Center}

\author[Idan Ginsburg \& Abraham Loeb]{Idan Ginsburg\thanks{E-mail:
iginsburg@cfa.harvard.edu} \& Abraham
Loeb\thanks{E-mail:aloeb@cfa.harvard.edu}\\ Harvard-Smithsonian Center for
Astrophysics, 60 Garden St., MS 51, Cambridge, MA 02138, USA\\}

\begin{document}
\maketitle

\begin{abstract}

The hypervelocity star SDSS J090745.0+024507 in the halo of the Milky Way
galaxy \citep{Brown:05} most likely originated from the breakup of a binary
star system by the central black hole, SgrA* \citep{Hills:88}. We examine
the fate of former binary companions to similar hypervelocity stars (HVSs)
by simulating 600 different binary orbits around SgrA* with a direct N-body
integration code.  For some orbital parameters, the binary breakup process
leads to HVSs with ejection velocities that are almost an order of magnitude
larger than the velocity observed for SDSS J090745.0+024507. The former
companion stars populate highly eccentric orbits which resemble the
observed orbits for some of the stars nearest to SgrA*.

\end{abstract}


\section{Introduction} \label{I}

Recently, the first hypervelocity star (HVS), SDSS J090745.0+024507, was
discovered in the Galactic halo \citep{Brown:05,Fuentes:05}.  This HVS is
located at a heliocentric distance of $\sim 71$ kpc and has radial velocity
853 $\pm$ 12 kms$^{-1}$.  Its velocity is over twice that needed to escape
the gravitational pull of the Milky Way. \citet{Hills:88} was the first to
recognize that a HVS might result from a close encounter between a tightly
bound binary star system and the black hole at the Galactic center, SgrA*.
\citet{Yu-Tremaine:03} refined Hills' argument and added that HVSs might
also be produced by three-body interactions between a star and a binary
black hole.  Because the existence of a second black hole in the Galactic
center \citep{Hansen:03} is only a hypothetical possibility
\citep{Scho:03}, we focus our discussion on the disruption of a tightly
bound binary by a single supermassive black hole (SMBH).

The Keplerian orbits of massive stars within $10^2$--$10^4$AU from the
Galactic center provide strong evidence for the existence of a central SMBH
with mass $\sim 4\times 10^6M_{\odot}$ (e.g. \citealt{Ghez:05};
\citealt{Reid-Brunthaler:04}; \citealt{Scho:03}).   Since binaries are
common in other star forming environments, it is only natural to explore
the interaction between SgrA* and nearby binaries.  Both \citet{Yu-Tremaine:03} 
and \citet{Gould-Quillen:03} describe instances where the tidal disruption of 
a binary by the SMBH leads to one star being ejected into close orbit around 
the black hole.  Thus, it is only natural to ask: {\it Is it possible that 
some of these stars are former companions of HVSs?}

In \S 2 we describe the
N-body code and simulation parameters that were adopted.  In $\S3$ we
discuss our numerical results for the origin of the HVSs, and in $\S4$ we
compare the calculated orbits of the bound companion stars to the observed
stellar orbits near the Galactic center. Our goal is not to cover the
entire phase space of possible binary orbits but rather to examine whether
some of the highly eccentric orbits of observed stars near SgrA* could have
resulted with a reasonable probability from the disruption of a stellar
binary.

\section{Computational Method} \label{CM}

In our study we have used the N-body code NBODY0 written by
\citet{Aarseth:99}, and presented in \citet{Binney-Tremaine}.  We have
tested NBODY0 against later versions of Aarseth's N-body codes (such as
``triple''), and found the results to be identical to within the required
precision.  We adopted a small value of $10^{-8}$ for the accuracy
parameter $\eta$, which determines the integration step through the
relation $dt=\sqrt{{\eta F}/(d^2F/dt^2)}$ where $dt$ is the timestep and
$F$ is the force.  The softening parameter, \emph{eps2}, which is used to
create the softened point-mass potential, was set to zero. We treat the
stars as point particles and ignore tidal and general relativistic effects
on their orbits, since these effects are small at the distance ($\sim
10$AU) where the binary is tidally disrupted by the SMBH.

We have set the mass of the SMBH to $M=4\times 10^6M_{\odot}$ and the mass
of each star to $m=3M_{\odot}$, comparable to the estimated mass of SDSS
J090745.0+024507 (Fuentes et al. 2005).  All runs start with the center of
the circular binary located 2000 AU ($=10^{-2}$pc) away from the SMBH along
the positive y-axis. This distance is comparable to the inner scale of the
observed distribution of stars around SgrA* (\citealt{Eckart-Genzel:97};
\citealt{Scho:03}; \citealt{Ghez:05}), allowing the remaining star to
populate this region after the ejection of its companion.  This radius is
also much larger than the binary size or the distance of closest approach
necessary to obtain the relevant ejection velocity of HVSs, making the
simulated orbits nearly parabolic.  We used the same initial distance for
all runs to make the comparison among them easier to interpret as we varied
the distance of closest approach to the SMBH or the relative positions of
the two stars within the binary.

We chose initial binary separations of $a=0.05$ or $0.1$AU because they
provide ejection velocities in the range of interest\footnote{Note that the
original ejection speed of SDSS J090745.0+024507 should have been higher
than its observed speed because of its deceleration in the Galactic
potential.} for the above parameters.  Significantly wider binaries would
give lower ejection velocities \citep{Gualandris:05}.  Much tighter
binaries would not be easily disrupted by the black hole, or may coalesce
to make a single star before interacting with the SMBH. The size of a main
sequence star of a few solar masses is $\sim 0.01$AU, and so binaries
tighter than $\sim 0.02$AU are precluded because the two stars will develop
a common envelope and eventually coalesce.  

In the Galactic disk, about half of all stars form in binaries or small
multiple systems (see e.g. \citealt{Duquennoy-Mayor:91}), with roughly
equal probability per logarithmic interval of separations,
$dP/d\ln(a)=const$ (e.g. \citealt{Abt:83}; \citealt{Heacox:98};
\citealt{Larson:03}). In the Galactic center environment, the maximum
binary separation is limited by the tidal force of SgrA* at the distance
$d$ where the binary is formed (for conditions that enable star formation
near the SMBH, see \citealt{M-Loeb:04}).  Since the mass of the black hole
is $\sim 10^6$ times larger than that of a star, this implies a maximum
binary separation less than $(10^{-6})^{1/3}=10^{-2}$ of the initial
distance $d$. For $d=2\times 10^3$AU, the upper limit on the binary
separation would be 20AU (or smaller if the tidal restriction applies
during the formation process of the binary). If we assume a constant
probability per $\ln(a)$ for $0.02<a<20$AU, then the probability of finding
a binary in the range of $a=0.05$--$0.1$AU is substantial, $\sim 10\%$.

We have found that the initial phase of the binary orbit plays a crucial
role in the outcome.  Therefore, we sampled cases with initial phase values
of: -75, -60, -45, -30, -15, 0, 15, 30, 45, 60, 75, and 90 degrees.  As
initial conditions, we gave the binary system no radial velocity but a
tangential velocity with an amplitude in the range between $5$ and
$500~{\rm km~s^{-1}}$ at the distance of 2000AU.  As described
analytically below, we expect no HVSs to be produced at larger tangential
velocities.  We ran 300 cases for each of the two binary separations, with
a total of 600 simulations.

\section{Origin of the Hypervelocity Star} \label{LC}

\begin{figure}
\begin{center}
\includegraphics[width=\columnwidth]{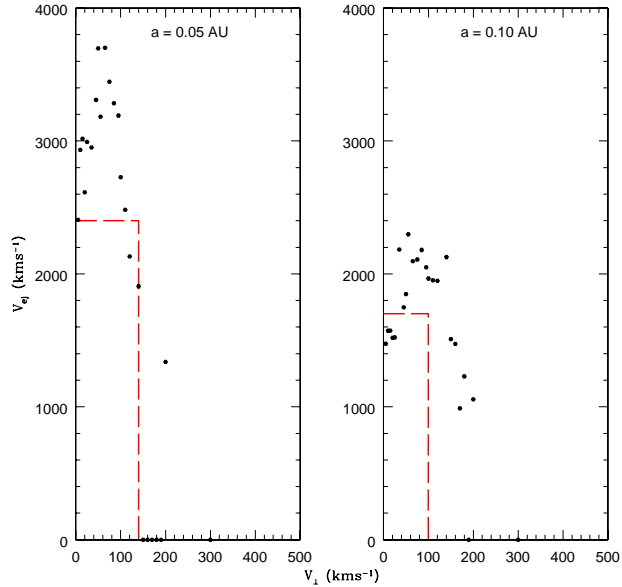}
\end{center}
\caption{Ejection speed $v_{\rm ej}$ of the unbound HVS as a function of
the initial tangential velocity $v_{\perp}$ of the binary at a distance of
$2000$ AU from SgrA*.  The dashed line indicates the analytic model of
Eqs. (\ref{eq:model}) and (\ref{eq:crit}).  Each point represents the
average ejection velocity for all HVSs at the given tangential velocity
(averaged over the initial orbital phase within the binary). The average
value for $v_{\rm ej}$ over all points agrees closely with our analytical
model.  For $a$ = 0.05 AU we get agreement within 16\%, and for $a$ = 0.10
AU we get agreement within 3\%.}
\label{evt}
\end{figure}

Given a binary system with stars of equal mass $m$ separated by a distance
$a$ and a SMBH of mass $M\gg m$ at a distance $b$ from the
binary, tidal disruption would occur if $b\la b_{\rm t}$ where
\begin{equation}
\frac{m}{a^3} \sim \frac{M}{{b^3_{\rm t}}}
\end{equation}
The distance of closest approach in the initial plunge of the binary
towards the SMBH can be obtained by angular momentum conservation from its
initial transverse speed $v_{\perp}$ at its initial distance from the SMBH,
$d$,
\begin{equation}
v_{\perp}d = \left(\frac{GM}{b}\right)^{1/2}b .
\end{equation}
The binary will be tidally disrupted if its initial transverse speed
is lower than some critical value,
\begin{equation}
v_\perp\la v_{\perp,\rm crit} \equiv {(GMa)^{1/2}\over d}\left({M\over
m}\right)^{1/6}= 10^2 {a_{-1}^{1/2} \over m_{0.5}^{1/6} d_{3.3}}
~{\rm {km~s^{-1}}},
\label{eq:crit}
\end{equation}
where $a_{-1}\equiv ({a}/{0.1~{\rm AU}})$, $d_{3.3}=(d/2000~{\rm AU})$,
$m_{0.5} \equiv (m/3M_{\odot})$, and we have adopted $M=4\times 10^6
M_{\odot}$.  For $v_\perp\la v_{\perp,\rm crit}$, one of the stars receives
sufficient kinetic energy to become unbound, while the second star is
kicked into a tighter orbit around the SMBH.  The ejection speed $v_{\rm
ej}$ of the unbound star can be obtained by considering the change in its
kinetic energy $\sim v\delta v$ as it acquires a velocity shift of order
the binary orbital speed $\delta v \sim \sqrt{Gm/a}$ during the disruption
process of the binary at a distance $\sim b_t$ from the SMBH when the
binary center-of-mass speed is $v\sim \sqrt{GM/b_t}$
\citep{Hills:88,Yu-Tremaine:03}. At later times, the binary stars separate
and move independently relative to the SMBH, each with its own orbital
energy.  For $v\la v_{\perp,\rm crit}$, we therefore expect
\begin{align}
v_{\rm ej} \sim \left[\left({\frac{Gm}{a}}\right)^{1/2}\left(
{\frac{GM}{b_{\rm t}}}\right)^{1/2}\right]^{1/2} \nonumber\\ = 
1.7 \times 10^3 
m^{1/3}_{0.5}a^{-1/2}_{-1} ~{\rm km~s^{-1}}.
\label{eq:model}
\end{align}

\begin{figure*}
\begin{center}
\begin{tabular}{ccc}
\includegraphics[width=0.52\textwidth]{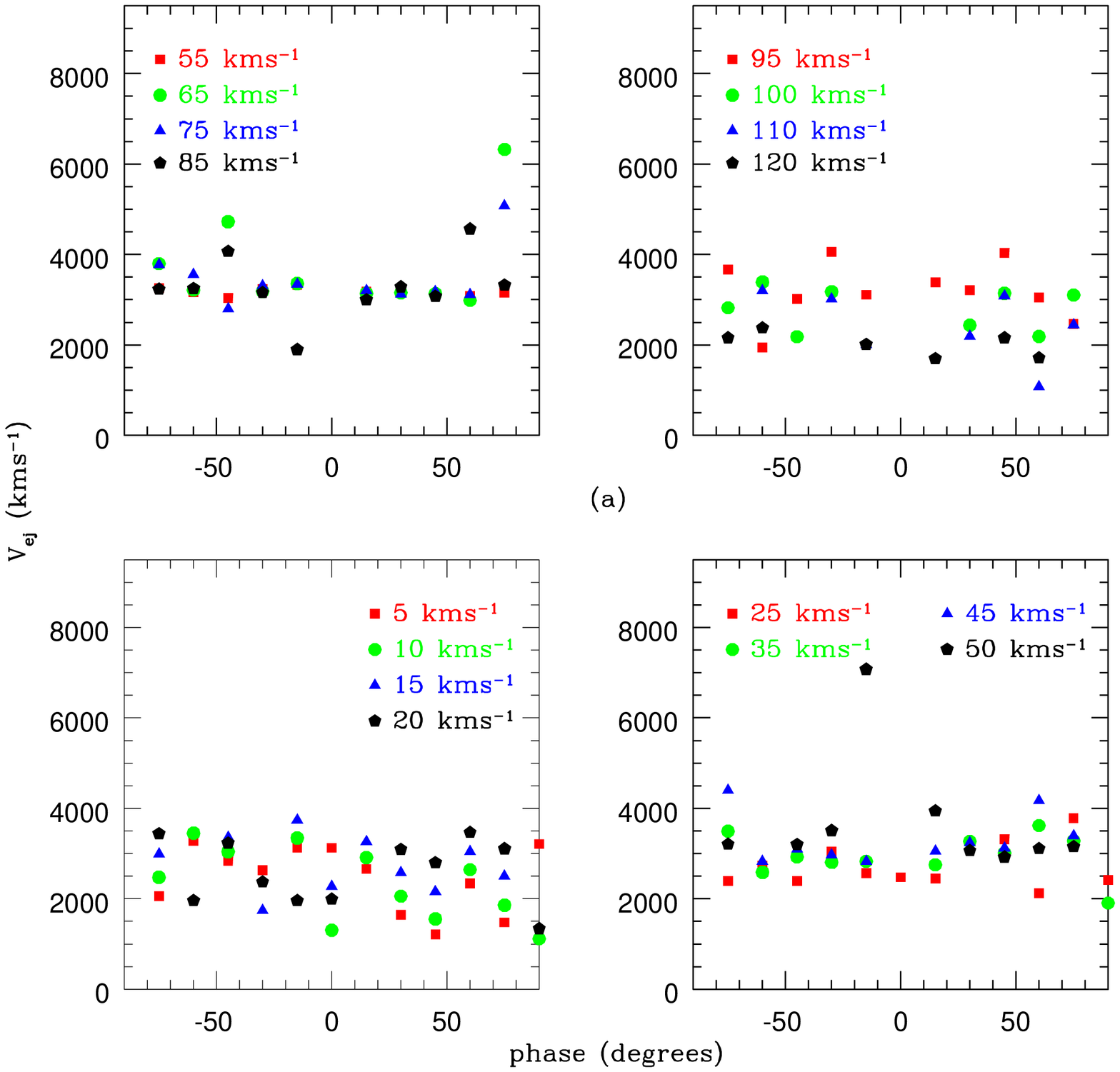}
&
\includegraphics[width=0.52\textwidth]{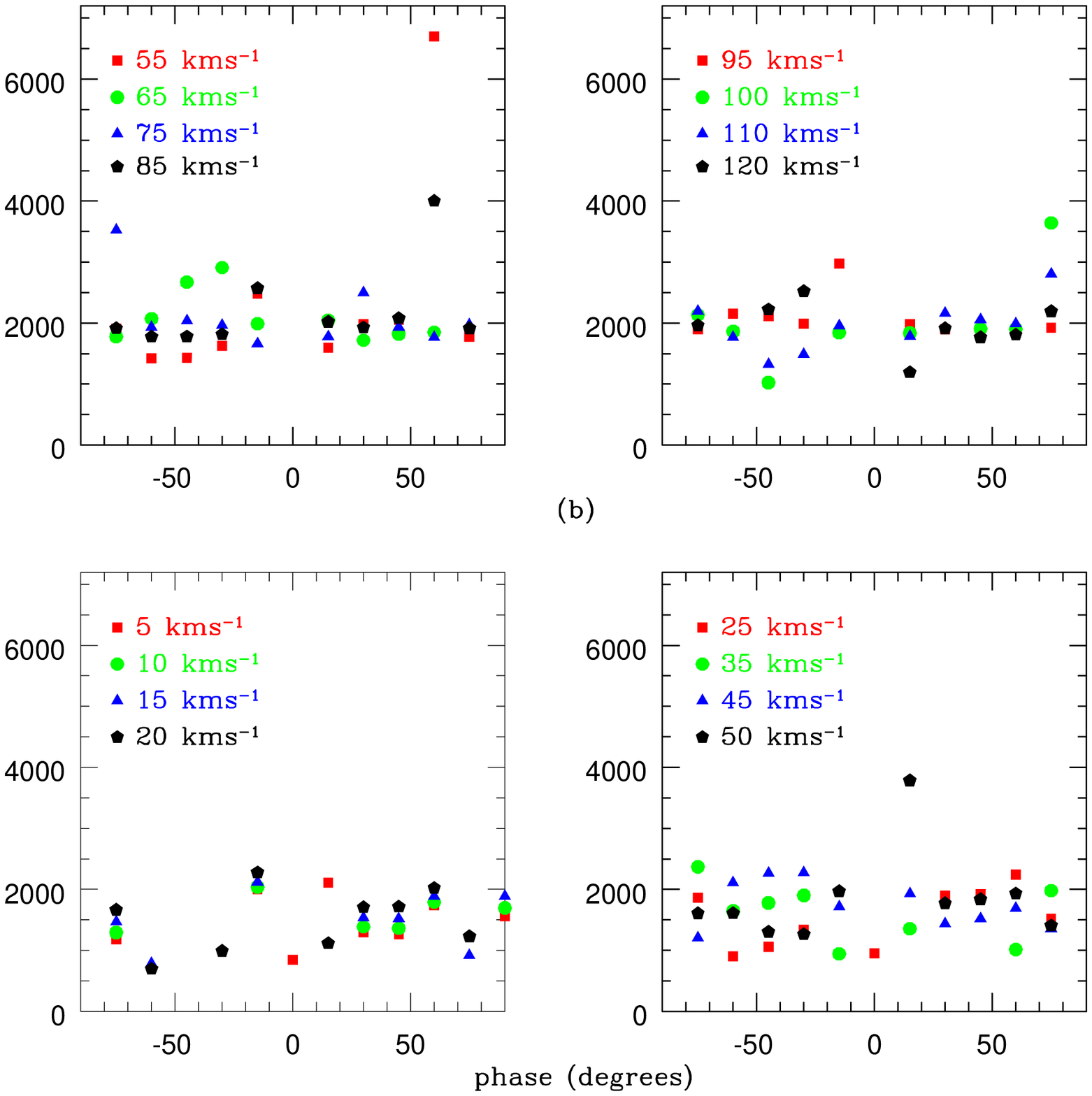}
&
\\
\end{tabular}
\end{center}
\caption{Ejection speed $v_{\rm ej}$ as a function of initial orbital phase
within the binary (in degrees).  The four cases in (a) refer to a binary
separation $a= 0.05$ AU.  The cases in (b) refer to $a = 0.1$ AU.  The
ejection velocity tends to lie within the range of 2000--$4000~{\rm
km~s^{-1}}$, but occasionally goes beyond it. The ejection speed is
sensitive to the orientation of the binary when it gets disrupted, which in
turn depends on the initial phase as well as the other (internal and bulk)
orbital parameters of the binary. The different point types refer to
different choices for the transverse speed of the binary relative to the
SMBH at their initial separation of 2000AU.}
\label{phase}
\end{figure*}

Figure \ref{evt} compares the above approximate model (dashed line) with
the results from our N-body simulations (points).  The expected values of
$v_{\rm ej}$ and $v_{\perp,\rm crit}$ (dashed line) and their dependence
on the binary separation $a$ in Eq. (\ref{eq:crit}), are consistent
with our numerical results. However, statistical variations exist. For
$v_{\perp}\la v_{\perp,\rm crit}$ the numerical runs show variations by up
to a factor of $\sim 2$ around the expected flat value of $v_{\rm ej}$ in
Eq. (\ref{eq:model}).  There are also exceptions of escaping stars with
$v_{\perp} \ga $ 200 kms$^{-1}$.

Hills (1988) concluded that the ejection speed of a HVS could reach a value
of $\sim 4000~{\rm km~s^{-1}}$.  Although the vast majority of our
simulated HVSs did not go beyond Hills' limit, there were a few exceptions 
(see Figure \ref{phase}).  Of our 600 runs, there were
307 stars that escaped the SMBH, and of those 12 had velocity $\geq$ 4000
kms$^{-1}$.  Furthermore, 3 had $v_{ej} \ga$ 6000 kms$^{-1}$, of which 
the largest had velocity $v_{ej} = 7073~{\rm km s^{-1}}$.  The likelihood of
observing such a HVS is remote.  As noted, HVSs with $v_{ej} \geq$ 4000 
kms$^{-1}$ rarely occur, and moreover a star with velocity 6000 kms$^{-1}$ 
would traverse the $\sim 200$ kpc scale of the Milky Way halo 
\citep{Wilkinson-Evans:99} in just 30 million years, reducing the likelihood 
for the observer to find it.

\begin{figure*}
\begin{center}
\includegraphics[width=\columnwidth]{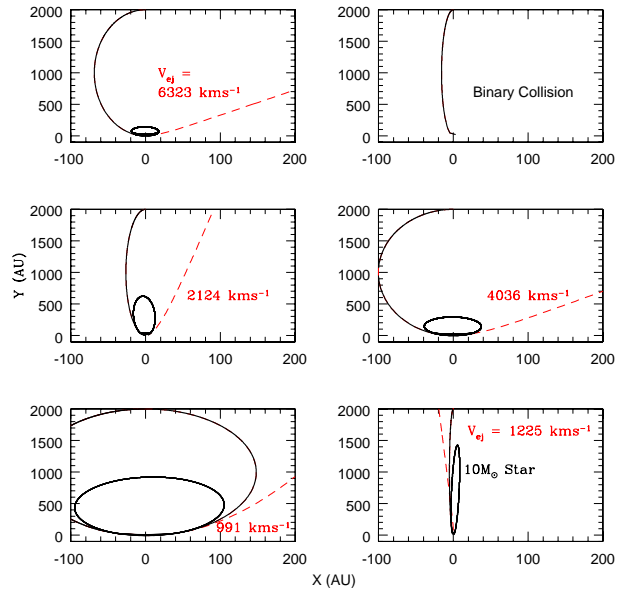}
\end{center}
\caption{Orbits of the former companions to HVSs.  The SMBH is located at
the origin, and the binary system is given an initial tangential velocity
$v_\perp$ along the negative x-axis at an initial distance of 2000 AU along
the positive y-axis.  The escape trajectory of the HVS is delineated by the
dashed line.  The binary is often disrupted upon its first revolution
around the black hole as seen in the bottom right and middle two plots.
However, sometimes it takes an extra revolution as seen in the other cases.
This is most likely caused by the changing phase of the binary at closest
approach.  Note that the x and y axis have different label scales. The top
right panel shows the orbit of the binary system until collision. The
bottom right panel shows a $10M_{\odot}$ star and a $3M_{\odot}$ companion
with initial orbital separation of $a$ = 0.2 AU.  In all other cases both
stars were $3M_{\odot}$ and had orbital separation $a$ = 0.05 AU.}
\label{comp}
\end{figure*}

\section{Fate of the Companion Star}
The orbits of a number of known stars around Sgr A* have been studied in
detail (see e.g. \citealt{Eckart-Genzel:97}; \citealt{Scho:03};
\citealt{Ghez:05}).  An intriguing question is how did some of these stars
obtain their highly eccentric orbits near the central SMBH.

Figure \ref{comp} shows the orbits of the companion stars for five HVSs
produced by our simulations.  All of the orbits have very high
eccentricities, ranging from $e= 0.966$ to $e= 0.999$.  Similar results
are obtained for other orbits, not shown in the figure.  Previously derived
constraints on HVS J090745.0+024507 \citep{Fuentes:05} indicated that its
former companion must have remained bound to SgrA* with an eccentricity
within the range $e$ = 0.97 to $e \sim$ 1, in agreement with our results.

\citet{Ghez:05} list stellar orbital parameters for seven stars near SgrA*.
All but one of the stars have high eccentricity, and in particular, S0-16
has a well defined eccentricity of 0.974 $\pm$ 0.016.  It has a semimajor
axis $a$ = 1680 $\pm$ 510 AU, and period $P = 36 \pm 17$ years.  Our runs
produce a smaller semimajor axis and period, with $a\sim 400$ AU and $P
\sim 4$ years, but these numbers can be changed by varying the stellar
masses (see the bottom right panel of Fig.~\ref{comp}) or the initial
binary eccentricity. \citealt{Scho:03} provide the eccentricity of six
stars; of particular interest are S14 ($=$SO-16) and S8 ($=$SO-4) with high
eccentricities of 0.97 and 0.98 respectively.  The estimated periods are 69
years and 342 years for semimajor axes of 3115 AU and 5982 AU,
respectively.  For S0-2, \citet{Ghez:05} provided a period of 14.53 years
and a semimajor axis of 919 AU, with an estimated eccentricity of 0.8670,
which is not within the range of our runs but close.
\citet{Gould-Quillen:03} suggested that the tidal disruption of a
massive-star binary could account for the orbit of star S0-2.  If so, a
star of order 100 $M_{\odot}$ must be the companion.  Although possible,
such an association is unlikely given the rarity of expected companion
stars with this mass (see \citealt{Kroupa:05}).  Stars on an eccentric
orbit may also be produced through an exchange reaction of a massive star
with a stellar-mass black hole on a tight orbit around SgrA*
\citep{Alexander-Livio:04}.

Owing to the flux limit inherent in infrared observations of the Galactic
center, the observed close-in stars near SgrA* are more massive than we
assumed in our analysis. For example, \citet{Ghez:03} estimated a mass of
$\sim 10M_\odot$ for S0-2.  An opposite selection effect applies to SDSS
J090745.0+024507, because stars with a mass $\ga 10M_\odot$ would not be
observable in the Galactic halo as their lifetime would be shorter than the
duration of their journey.  The scaling in Eq.  (\ref{eq:model}) implies
only a modest change in $v_{\rm ej}$ for $10M_\odot$ stars.  As an example,
the bottom right panel of Figure \ref{comp} shows the outcome of the
disruption process of a tight ($a=0.2$AU) binary containing stars of
$10M_{\odot}$ and $3M_{\odot}$. The $3M_\odot$ star is ejected while the
$10M_\odot$ companion remains in a highly eccentric orbit with $e$ = 0.999.
The orbital semimajor axis is 700 AU, and the period is 9.3 years.  A
different run with a $0.4$AU binary starting at 4000 AU with a transverse
speed of $5~{\rm km~s^{-1}}$ left the $10M_\odot$ companion with an orbital
semimajor axis of 1430 AU and an eccentricity of $e=0.99$, closer to the
observed parameters of S14.

{\it What is the likelihood for a collision between the stars as a result
of the kick they acquire from their interaction with the SMBH?}  In our
runs, the binary was taken to lie in the same plane as its orbit around the
SMBH.  Assuming that the impulsive kick is given by the SMBH towards a
random direction within this plane, the probability for a collision in a
case that otherwise would have produced a HVS is four time the radius of a
star (which is $\sim 0.01$ AU for a $3M_{\odot}$ star) divided by the
circumference of a circle with a radius equal to the binary separation.
For $a=0.05$AU and $0.1$ AU this would imply a collision probability of
12.73\% and 6.37\%, respectively.  Our runs gave consistent statistical
results, with a collision fraction among HVS orbits of $\sim 7.7\pm2.1\%$
for $a= 0.05$AU and $5.6\pm 1.9\%$ for $a= 0.1$AU.  
The likelihood for a collision is expected to be smaller in the more general
case when the binary lies in a different plane than its orbit around the
SMBH. Averaging over all random orientations, the collision probability is
$[\pi(0.02 AU)^2/4\pi(0.05 AU)^2]= 4\%$ for $a= 0.05$AU and 1\% for $a=
0.1$AU. The small likelihood introduces only a minor correction to our
earlier statistical results (which were obtained by approximating the stars
as point-like particles).  An example of an orbit that leads to a collision
is shown in the top-right panel of Figure \ref{comp}.  The two stars would
not merge as a result of the collision if their relative speed 
significantly exceeds the escape speed from their surface ($\sim 500~{\rm
km~s^{-1}}$).  In the example shown, the relative speed of the stars at
impact was $666~{\rm km~s^{-1}}$.  Grazing-incidence collisions, which are
more probable than head-on collisions, may lead to HVSs which are rapidly
spinning \citep{Alexander-Kumar:01}. 

Finally, we note that for the typical impact parameter that leads to the
break-up of the binary with $a\sim 0.05$AU by the black hole (Eq. 1), the
tidal force on a star is less than a few percent of the gravitational force
that binds the star; however, some rare encounters (with $a$ replaced by
the stellar radius in Eq. 1) may lead to the tidal disruption of the stars.

\section{Conclusions} \label{Im}
Our N-body simulations indicate that tight binaries with separations
$0.05$--$0.1$AU which approach within a distance $\la 10$AU from SgrA*
could produce HVSs with velocities almost an order of magnitude greater
than the observed velocity of HVS SDSS J090745.0+024507.  The orientation
of the binary at closest approach plays an important role in determining
whether the binary is tidally disrupted by the SMBH and what is the
eventual ejection velocity of the unbound star in that case (see
Fig. \ref{phase}).  The phase sensitivity originates from the fact that the
crossing time of the distance at which tidal disruption occurs,
$(GM/b^3)^{-1/2}$ is shorter than the binary orbital time
${\sqrt{2}}*pi*(Gm/a^3)^{-1/2}$.  Numerically, we have not been able to
identify a simple trend for the phase angle at the breakup radius that
would appear more organized than the results in Figure \ref{phase}.

The former companion star to a HVS is typically kicked into a highly
eccentric orbit with an eccentricity $\geq$ 0.966 (see Fig. \ref{comp}).
The resulting eccentricity is similar to that observed for a number of
stars near the Galactic center such as 
S8 ($e=0.98$) and S14 ($e=0.97$), suggesting a possible binary origin for
these stars.

Surveys of HVSs in the halo of the Milky Way galaxy select moderate-mass
($m\la 5M_\odot$) stars with lifetimes longer than their travel times ($\ga
10^8$ years) while infrared surveys of the vicinity of SgrA* select for
bright massive stars ($m\ga 10M_\odot$) with short lifetimes ($\la
10^7$years). The findings of existing surveys suggest that both populations
of stars co-exist.  Future extensions of this work may examine a larger set
of binaries with various stellar masses and internal orbital eccentricity.

\section*{Acknowledgments}

We thank Sverre Aarseth, Avery Broderick, Warren Brown, Suvendra Dutta,
Mark Reid, and Rosanne Di Stefano for useful discussions, and the referee
Michael Sipior for helpful suggestions.  This work was supported in part by
Harvard university funds and by NASA grants NAG 5-1329 and NNG05GH54G.

\bibliographystyle{mn2e.bst}
\bibliography{Paper.bib}

\bsp

\end{document}